\documentclass{article}
\usepackage[preprint]{neurips_2026}

\usepackage{amsmath, amssymb, graphicx, float, wrapfig, enumitem}
\usepackage{xcolor}
\definecolor{amethyst}{rgb}{0.6, 0.4, 0.8}
\definecolor{olivegreen}{rgb}{0.33, 0.42, 0.18}

\definecolor{ultramarine}{rgb}{0.07, 0.04, 0.56}

\title{Learning Scattering Amplitudes with Transformer Reinforcement Learning }

\author{
  Philip Velie\textsuperscript{1}
  \qquad
  Piyush Jha\textsuperscript{2,3}
  \qquad
  Vijay Ganesh\textsuperscript{2}
  \qquad
  Aishik Ghosh\textsuperscript{3,4}
  \\[1ex]
  \textsuperscript{1}Institute for Gravitation and the Cosmos, Department of Physics,
Pennsylvania State University \\
  \textsuperscript{2}School of Computer Science, Georgia Institute of Technology, USA \\
  \textsuperscript{3}School of Physics, Georgia Institute of Technology, USA \\
  \textsuperscript{4}Lawrence Berkeley National Laboratory, USA \\
}

\begin{document}

\maketitle

\begin{abstract}

We introduce a transformer reinforcement learning algorithm that learns to solve high loop-level scattering amplitudes in planar $\mathcal{N}=4$ Super Yang-Mills theory. Our algorithm improves on previous transformer-only results \cite{Cai:2024znx} by the incorporation of previously derived symmetries and relationships into the learning algorithm. This results in a greatly decreased fraction of the solution that needs to be known a priori to solve the entire problem.
An additional benefit is that our algorithm also ensures that every output 
 obeys the set of known relationships. Rather than predicting all coefficients independently, the model proposes assignments that are propagated through exact linear relations, while MCTS searches over assignments when propagation alone is insufficient. This is crucial to the generalization of machine learning approaches to higher loops, as without this, there is no way to overcome the factorially scaling of state sizes and compare to results derived via other methods. Using the symbology representation of the form factor, we frame the problem as learning a mapping between discrete sequences and integer coefficients. 

\end{abstract}



\section{Introduction}

Scattering amplitudes are central objects in quantum field theory (QFT), encoding transition probabilities between asymptotic particle states. Despite their fundamental importance, the difficulty of traditional perturbative computations based on Feynman diagrams scales factorially with both loop order and the number of external particles, motivating the development of alternative frameworks. Planar $\mathcal{N}=4$ Super Yang-Mills (SYM) theory has emerged as a key testing ground for such developments due to its high degree of symmetry and simplified analytic structure. In this context, the symbol bootstrap program has enabled the computation of amplitudes at high loop orders by exploiting algebraic and physical constraints.

Recent work \cite{Cai:2025atc}, \cite{Cai:2024znx} has shown that the symbolic structure of amplitudes can be treated as a sequence modeling problem, making it amenable to modern machine learning approaches such as transformers. The shortcomings of this method are that it requires a vast majority of the final answer to be known so that it may be used as training data. Furthermore, if one only greedily samples the resulting probability distribution, the outcome is almost guaranteed to violate the set of known relationships. This suggests the need for additional structure. We extend this work with additional reinforcement learning, which allows us to leverage the many symmetries of $\mathcal{N}=4$ SYM as inputs and use fewer labeled pretraining data points. These relationships can deterministically infer many coefficients from a partial assignment, but propagation eventually reaches a fixed point with unresolved coefficients remaining. We therefore formulate reconstruction as a sequential search problem: a transformer proposes coefficient assignments, exact relations propagate their consequences, and MCTS explores alternatives when propagation alone is insufficient. This allows us to use substantially fewer labeled coefficients while ensuring consistency with the imposed relations, while retaining the proven advantage of carrying a $p(\text{coefficient}|\text{word})$ across loop levels \cite{Cai:2024znx}.
\begin{figure}[H]
    \centering
    \includegraphics[width=0.99\linewidth]{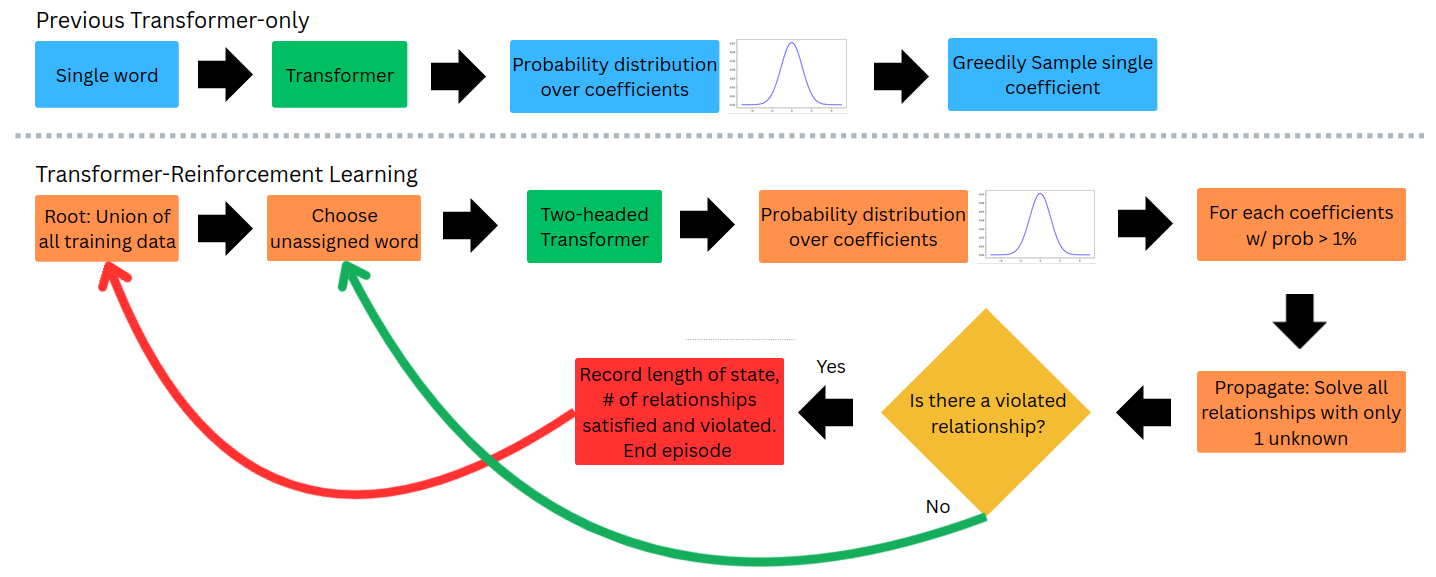}
    \caption{Comparison of transformer-only and transformer reinforcement learning approaches.}
    \label{fig:alg_diag}
\end{figure}

\section{Physics Background}



\paragraph{Three-Gluon Form Factor in $\mathcal{N}=4$ Super Yang Mills: }

Planar $\mathcal{N}=4$ Super Yang Mills (SYM) has the largest number of symmetries that a spin-1 gauge theory can have, but the most important for this analysis is that all of the Feynman integrals feature polylogarithms with maximum weight, $2L$. We consider the three-gluon form factor of the chiral stress-tensor multiplet operator $\mathrm{tr}(F_+^2)$ in planar $\mathcal{N}=4$ SYM. This observable is closely related to the $3g \rightarrow H$ amplitude in QCD in the large top-mass limit and provides a minimal nontrivial setting for studying amplitude structure.




\paragraph{Symbology: }



Amplitudes in $\mathcal{N}=4$ SYM, are often constrained to be nested series of polylogarithms. We can define a map, $S$, called the \textit{Symbol} \cite{Goncharov:2010jf}, that makes the manipulation of these integrals more explicit. This symbol satisfies the usual log properties, which we can use to construct a concise ``alphabet'' of rational functions of (dual) momenta. Different choices of alphabets can result in more compact symbols. For the three-gluon form factor in $\mathcal{N}=4$ SYM, the symbol alphabet consists of six letters $\{a,b,c,d,e,f\}$ where each letter is a combination momentum whose exact definition isn't relevant for this analysis. The $L$-loop symbol takes the form:
\begin{equation}
\mathcal{S}[F^{(L)}] =
\sum_{l_{i_1}, \dots, l_{i_{2L}}}
C_{l_{i_1}, \dots, l_{i_{2L}}}
\, l_{i_1} \otimes \cdots \otimes l_{i_{2L}},
\end{equation}
where the coefficients $C$ are integers. Thus, the amplitude problem reduces to determining integer-valued coefficients over sequences, called "words", of length $2L$ drawn from a six-letter alphabet. The naive total number of possible words grows as $6^{2L}$, leading to an exponential combinatorial explosion.










\paragraph{Adjacency Constraints and Sparsity: }

Despite the large combinatorial space, the symbol space is extremely sparse. Only a tiny fraction of words have nonzero coefficients, and this sparsity is largely governed by adjacency constraints. The term "legal" will be used to describe words that satisfy these properties. 
\begin{enumerate}[itemsep=0pt]
    \item Forbidden adjacent pairs include $\{a,d\}$, $\{b,e\}$, $\{c,f\}$, $\{d,e\}$, $\{e,f\}$, $\{d,f\}$.
    \item Words must be constructed with an alternating $\{(a,b,c)(d,e,f)(a,b,c)...(d,e,f)\}$ structure, where $(a,b,c)$ represents any string of letters ``a", ``b", or ``c", $(d,e,f)$ represents a string of letters ``d", ``e", ``f" where only one of these characters is repeated. 
\end{enumerate}

 
\paragraph{Integrability and Linear Relations: }
In addition to the adjacency constraints, the symbol must satisfy a set of relationships. Some of these relationships have exact physical meaning, such as integrability conditions or causality, while others do not. These relationships were first enumerated in \cite{Dixon:2022rse} and then expanded upon in \cite{Cai:2025atc}. These constraints lead to large systems of linear equations among coefficients. It is this novel integration of this set of relationships into our pipeline that allows us to derive the L=5 results with a drastically smaller pretraining set. For the purpose of preventing the transformer from emitting the trivial solution of ``all coefficients are 0", we add a smaller set of constructed non-homogeneous relationships. These relationships are constructed from the all-loop relationships from \cite{Cai:2025atc}. 










\section{Methodology}

\paragraph{Dataset preparation: }

In \cite{Dixon:2022rse}, the authors calculated the symbol for the $3g\rightarrow H$ form factor up to L = 8. Via the all-loop relationships, we know what the coefficients of a small percentage of the symbol terms should be. To emulate this, we take a fixed percentage of the symbol as the known pretraining dataset. For the result shown below, we first start by taking $50\%$ of the dataset as known before decreasing it further. Words are tokenized as sequences over the alphabet $\{a,b,c,d,e,f\}$, while coefficients are represented in a fixed integer encoding. Multiple tokenization schemes were tried, such as per-key tokenization, but the authors had the best success with the per-character tokenization. Since the size of the states scales by roughly $10^L$, we need to employ two different techniques to allow this method to tackle larger and larger loop levels.

\paragraph{Minimal suffix representation: }
\label{sec:rep}

Given that we have a set of relationships that only work at the end of words, we can employ the following technique to construct a compact representation of the states using our constructed set of relationships \cite{Dixon:2022rse}.

\begin{enumerate}[itemsep=0pt]
    \item To begin creating the Kth compact representation, construct all suffixes of length K. These suffixes are words that only need to end in ``def" and contain no illegal chunks 
    \item For each suffix, construct all relationships that employ the last chunk of the word 
    \item Assign an arbitrary order to the suffixes to construct a (sparse) matrix of relationships 
    \item Put said matrix in RREF and find the independent variables
    \item Finally, reduce the independent basis by removing all variables that are equivalent up to a $C_3$ symmetry
\end{enumerate}

The space of all possible words is now all possible legal combinations of prefixes (defined as words of length 2l-k that start with ``abc", and contain no illegal chunks) and suffixes. Each element of the independent basis is added to the alphabet, and each suffix is replaced by a representative token. This is the trade-off of this representation: smaller state sizes but larger token alphabet. This representation not only decreases the overall number of terms in the state but also decreases the size of each word to $2l-k+1$ tokens in length. 




\section{Implementation}
Here, we treat the state as the set of assigned word-coefficient pairs, and an action as assigning a coefficient to a yet unassigned word. The overarching theme of this algorithm is to select a word with an unassigned coefficient, have the transformer propose possible coefficients, propagate the consequences of assigning the coefficient via the relationships, and finally have the MCTS choose among the remaining non-violating branches. A diagramatic overview of this process is given in Figure \ref{fig:alg_diag}.





\paragraph{Pretraining: }
As opposed to the original Alphazero implementation \cite{silver2017masteringchessshogiselfplay}, we must first pretrain the transformer. Since the coefficients to our symbol words can be any $x \in \mathbb{Z}$, we use our pretrained transformer as a way to constrain our search space. However, this algorithm is inspired by \cite{Jha:2026qhs} as we are using a two-headed transformer to guide a Monte-Carlo Tree Search (MCTS). 

To begin pretraining, we create a set of roots where each root is a random subset of the a priori known subset. Each root is then propagated using the set of relationships to a fixed point. For each root, we then find the word that participates in the highest number of relationships that only have 2 unknowns. If the word is contained in the known subset, construct a set that contains both the correct value as well as random integers. Each integer in this set will be a child node. These steps are then repeated for each child node until a node budget is exhausted. The policy head of the transformer is trained with normal supervised learning between the most likely coefficient from its probability distribution. The value head, however, is trained based on the MSE between its predicted length of the terminal path and its actual length. The value head provides a long-range approximation to the short-range metric of the number of relationships that are satisfied or violated by a given state. 

\textbf{Pretraining and single-player Alphazero*:}
%
%
%
%
%
%
To begin the Alphazero phase of the algorithm, we then take a single starting point: the entirety of the known subset. Following this, we then find the top ten words that participate in the highest number of relationships that only have 2 unknowns. Passing one of these words at random to the pre-trained transformer, we extract an ensemble of probability distributions of logits from the policy head. We then collect these into a single probability distribution over potential coefficients. Each potential coefficient over a cut-off threshold and their propagated terms becomes a potential child node. MCTS begins over these child nodes with the value head's basis. We repeat these steps until the state is solved or until there is a violation. A violation is treated as a ``game over" signal.







\section{Results and Conclusion}

We applied this algorithm on the $L = 5$ symbol with the $K = 5$ representation, which consists of a total of 12,543 words. We used a known fraction of $f \in \{5,10,15,20,25,30,40,50\}\%$. At every fraction that we tested, the search assigns a coefficient to all 12,543 keys that satisfies every relationship and agrees with \cite{Dixon:2022rse}, up to an additional cyclic transformation. This compares with the $97\%$ of the symbols required for training in the transformer-only approach for L=6 (the L=5 result was not reported)\cite{Cai:2024znx}. Figure 2 demonstrates this entire ablation study. Propagating the seed accounts for roughly $70\%$ of all words before any work has to be done by the MCTS. Each trajectory demonstrates two regimes. The first regime is where propagation greatly assists the predictions by the transformer: contributing roughly 3 additional word-coefficient pairs for every one from the transformer. The second regime represents the roughly linear regime where most of the relationships with only 2 unknowns are exhausted, so more predictions have to be made to get a single propagated word-coefficient pair. This is the regime where the majority of the work is done by learned priors. 
\begin{figure}[H]
    \centering
    \includegraphics[width=0.80\linewidth]{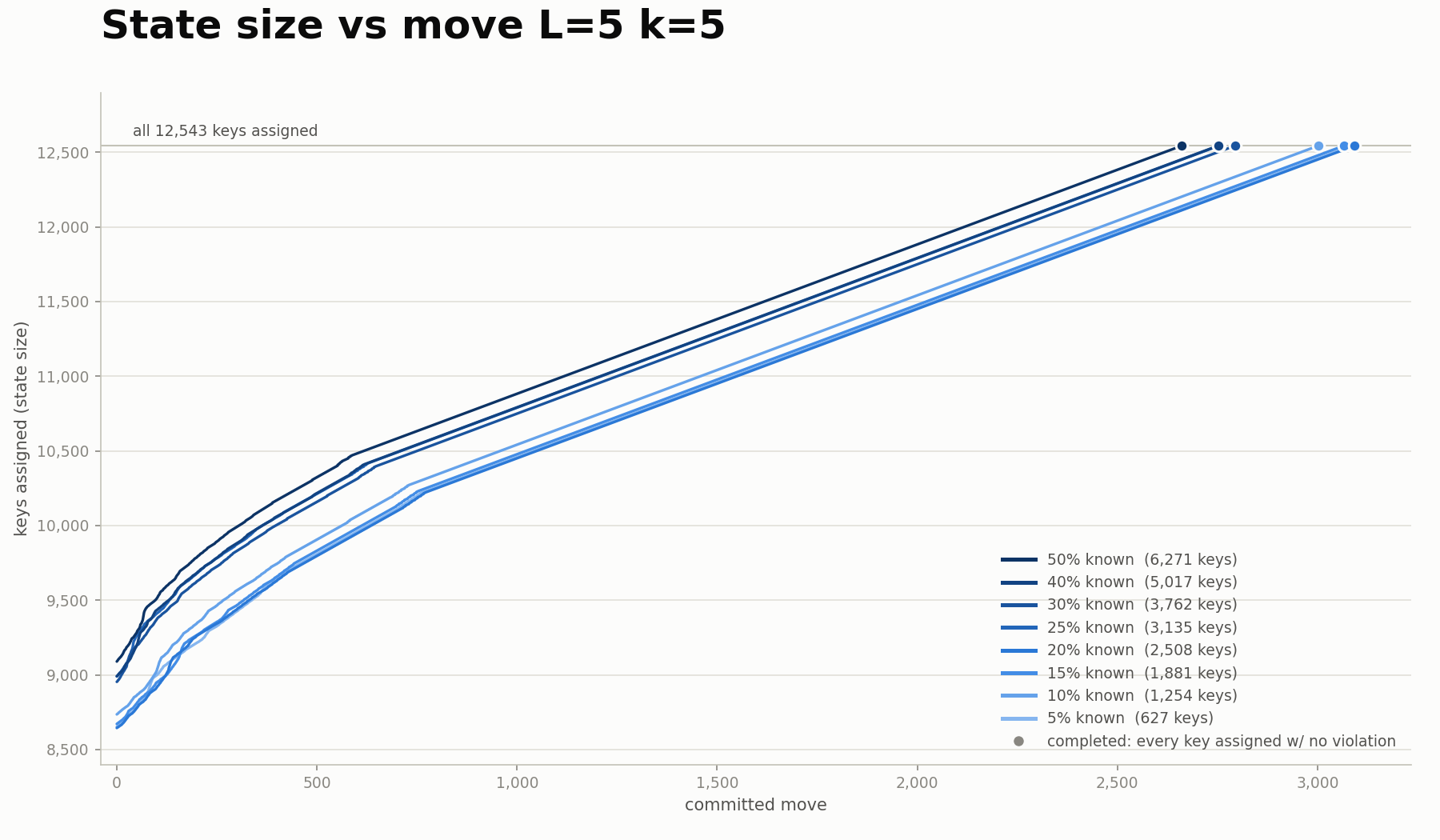}
    \caption{Ablation study on the $L=5$, $K=5$ symbol for known fractions $f$
from 5\% to 50\% (darker = larger $f$)}
    \label{fig:placeholder}
\end{figure}
This demonstrates that the pretrained transformer combined with propagation granted by the symmetries of $\mathcal{N}= 4$ SYM and \cite{Cai:2025atc}, along with the Monte-Carlo Tree Search, reconstructs the complete $L = 5$ symbol of the $3g\rightarrow H$ with as little as $5\%$ of its coefficients. By construction, any violation ends the MCTS search path; every completed output of this algorithm satisfies all these imposed relationships. This is not guaranteed in the transformer-only approach as there is no additional structure to constrain its output.    

\paragraph{Note added:} This work was submitted to ML4PS on 19th September 2026. Anthropic released their results on September 25th with loop 9. Our approach uses significantly less computational power but has yet to be demonstrated on loop 9, and will be considered in follow-up work.

\section*{Acknowledgments}
We thank Lance Dixon for fruitful discussions. This research used computing resources allocated to AG’s NESAP `AI Physicist' project at the National Energy Research Scientific Computing Center (NERSC), a U.S. Department of Energy Office of Science User Facility located at Lawrence Berkeley National Laboratory.

\bibliographystyle{plain}
\bibliography{amp}

\end{document}